# Survey on Architectural Attacks:
# A Unified Classification and Attack Model

Tara Ghasempouri, Jaan Raik

Tallinn University of Technology
Department of Computer Systems
tara.ghasempouri@taltech.ee

Cezar Reinbrecht, Said Hamdioui, Mottaqiallah Taouil

Delft University of Technology
Faculty of EE, Mathematics and CS
m.taouil@tudelft.nl

*Abstract*—According to the World Economic Forum, cyberattacks are considered as one of the most important sources of risk to companies and institutions worldwide. Attacks can target the network, software, and/or hardware. During the past years, much knowledge has been developed to understand and mitigate cyberattacks. However, new threats have appeared in recent years regarding software attacks that exploit hardware vulnerabilities. We define these attacks as architectural attacks. Today, both industry and academy have only limited comprehension of architectural attacks, which represents a critical issue for the design of future systems. To this end, this work proposes a new taxonomy, a new attack model, and a complete survey of existing architectural attacks. As a result, our study provides the tools to understand the Architectural Attacks deeply and start building better designs as well as protection mechanisms.

## I. INTRODUCTION

The importance of cybersecurity grows every year. Future projections foresee a total market growth from 155.83 billion US dollars in 2022 to 376.32 billion US dollars by 2029 [1]. The main reason relies on deploying new technologies like Internet-of-Things (IoT) and 5G communication. These technologies significantly increase the number of devices and their connectivity [2]. As a result, such conditions create new opportunities for cyberattacks. There are many types of cyberattacks [3]. They can be organized based on the target system, being Network, Software and Hardware [4]. On the otherhand, the attacks can also be classified based on the attack vector (i.e., the actor responsible for performing the attack), which are software (or logical) and hardware (or physical) [5].Combining both target and vector reveals the main types of threats present in the cyberattack field. For example, a virusis a software attack that targets another piece of software.A Distributed-Denial-of-Service (DDoS) [6] is a software attack that targets the network. Furthermore, a fault injection attack [7] is a hardware attack that targets the hardware. Most of the presented examples are known in academia and industry for many years, where mature protection mechanismsalready exist. However, in the last years, new attacks where software attacks target the hardware have emerged rapidly.We define such a new segment of attacks as *Architectural Attacks* (ArchA). These attacks can compromise the entiresystem by exploiting a hardware vulnerability while being operated by software. As software, they can be realized even remotely. Additionally, the continuous increase of architectural complexity and connectivity in current and future chips show a clear trend that Architectural Attacks will only appear increasingly. Therefore, there is an urgent need to understand how they behave, what they have in common and how they work to build intelligent and efficient countermeasures.

There are two main steps towards the comprehension of Architectural Attacks (ArchA). The first step is to define an appropriate taxonomy. Such taxonomy can put in evidence the main features that define each type of attack. Currently, there are only proposed classifications for a subset of the ArchA. Cache-based attacks are the most studied group of attacks having many proposed classifications [8–10]. More recently, transient execution attacks have gained popularity and some classifications have also been proposed [11]. Other schemes have also been proposed to classify software attacks that exploit hardware based on their level of sharing in the system [12, 13]. However, no scheme has successfullyput together all ArchA into the same taxonomy yet. The second step required to understand ArchA relies on having a representative attack model. An attack model is a formal description that organizes any attack as a sequence of generic actions. Consequently, such models define the patterns inside each attack, revealing its *backbone*. Similarly to the taxonomy problem, most attack models describe only cache attacks [14, 15]. Other works tried to describe how attacks behave, but they lack the required formalism to be considered as an attack model [16–18]. Therefore, it is clear that current taxonomies and attack models can not organize and describe the Architectural Attacks because they can only refer to a subset of attacks.

This paper proposes a new taxonomy and attack model to organize and describe the Architecture Attacks. As a result, we present a survey of the existing attacks showing how each one fits into our classifications and modelling. Hence, our main contributions are the following:

- Proposal of a new taxonomy for Architectural Attacks using three metrics: what, where and how.
- Proposal of a new attack model for Architectural Attacks which contains five steps: Setup, Trigger, Operation, Retrieve and Evaluation .

- A survey on the existing Architectural Attacks, framing them into the proposed classification and model.

The remainder of the paper is organized as follows. Section II describes the literature review systematically. Section III summarizes the related works that proposed taxonomies and attack models. Next, Section V introduces our proposed taxonomy. After that, Section IV describes our proposed attack model. Afterwards, we present a survey of the existing attacks in three different section; Section VI regarding IP Attacks, Section VII regarding Functionality attacks, and Section VIII regarding Data Attacks. Section IX provides a discussion on the outlooks and lastly Section X concludes the paper.

## II. RESEARCH METHODOLOGY

The literature covers a wide range of attacks that can compromise the entire system by exploiting a hardware component while being operated by software, i.e. Architectural Attacks (ArchA). The used methodology is based on Kitchenham and Charter's methodology [19], shown in Fig. 1. In the following the actions in each step of Kitchenham and Charter's methodology are described.

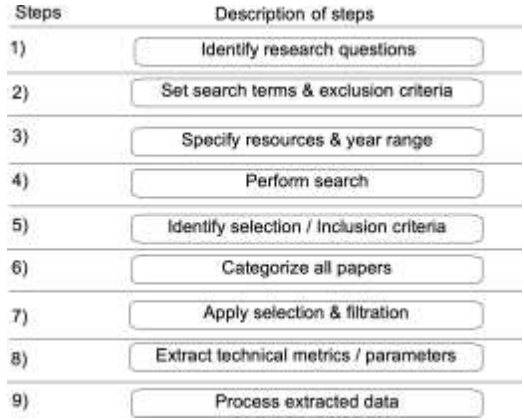

Fig. 1: Steps of Kitchenham and Charter's methodology.

*Step 1: Identify research questions*: The objective of this survey is set to answer research questions concerning ArchA, as follows:
- RQ1: What is a general taxonomy for ArchA that can represent main features of each type of attacks? This question is an essential since no scheme has successfully put all ArchA together in a same taxonomy.
- RQ2: What is a formal description to model ArchA? This question is important since describing attack models in a formal sequence of actions can clearly represent the patterns inside the attacks.

*Step 2: Set Search Terms & Exclusion Criteria*: The second stage of conducting this review is to identify the search terms such as keywords used in searching and collecting relevant papers in order to answer the research questions.

The below keywords with the following strategy are used: First we started by generic keywords, i.e., 'attacks', 'hardware', and 'software' since, the focus of this survey is on attacks which target hardware by manipulating software. After, we added detailed keywords such as 'processors', 'cache memory' and communication channels'

Exclusion criteria are as follows:
- Non peer reviewed sources such as online articles.
- Attack mitigation papers that do not incorporate the attack mechanism itself.
- Articles that discusses attacks at only software level.

*Step 3: Specify Resources & Year Range*: The following digital libraries are used in order to get all the related articles: IEEE Explorer, Google Scholar, ACM Digital Library, Elsevier, and Springer. All articles between 2011 and 2021 have been considered.

*Step 4: Perform Search*: Initially, a total of 2065 papers were found based on the initial search terms mentioned earlier. Then, the papers were filtered down to more focused categories. How these papers are filtered down is described next.

*Step 5: Identify selection / Inclusion criteria*: After analyzing articles found we noticed that these attacks are mainly performing on three domains, i.e., 'Processor', 'Cache memory' and 'Communication channel', thus we narrowed the search space by the corresponding keywords.

*Step 6: Categorize all papers*: Table I represents the number of articles for each search, based on keywords defined in previous steps.

TABLE I: Search result count of respective keyword combinations.

| Search keywords | Number of articles |
|---|---|
| 'hardware', 'software', 'attack' | 2065 |
| 'hardware', 'attack, 'processor' | 520 |
| 'hardware', 'attack', 'cache memory' | 104 |
| 'hardware', 'attack, communication' | 1604 |

*Step 7: Apply selection and filtration*: The selection and filtration process are explained below:
- Remove duplicated or multiple versions of the same articles.
- Apply inclusion and exclusion criteria to avoid any irrelevant papers.
- Remove review papers from the collected papers.
- Apply quality assessment rules to include qualified papers (quality is determined based on the parameters mentioned in the next step) that best address the research questions.
- Look for additional related papers using the reference lists of the collected papers and repeat the same process again.

*Step 8: Extract technical metrics / parameters:* Applying some parameters such as quality assessment rules (QARs) is the step used to identify the list of papers that are consideredin this review. The QARs are important to ensure the proper evaluation of the research papers' quality. Therefore, 10 QARs are specified, each worth 1 mark out of 10. The score of each QAR can be 1 and 0. 1 refers to "answered" while 0 refers to "not answered". The selected article should get the score 10 out of 10 to be selected, otherwise it is excluded. The QARs are listed below:

- QAR1: Are the research objectives identified clearly?
- QAR2: Are the techniques of the implementation well defined and deliberated?
- QAR3: Is the platform (in this case, processor, cache memory and communication protocol) that ArchA going to apply on it clearly defined?
- QAR4: Does the paper include practical experiments of the proposed technique?
- QAR5: Are the experiments well designed and justified?
- QAR6: Is the proposed algorithm applied on a hardware component?
- QAR7: Is the method effectiveness reported?
- QAR8: Is the proposed ArchA model design compared to others?
- QAR9: Are the methods used to analyse the results appropriate?
- QAR10: Does the overall study contribute significantly to the area of research?

Overall, after performing Step 7 and Step 8, we identified 95 useful papers.

*Step 9: Process extracted data:* In this step, the objective is to analyze the final list of papers in order to extract the needed information, i.e., provide answers to the given research questions. First, general details are extracted from each paper such as the paper number, paper title, publication year, and publication type. Then, more specific information is sought, such as the algorithm models and the category of hardware component, whether it addresses cache memory, processor, communication channel or a combination of them. Finally, any details directly related to the research questions are pursued. Table I shows the number of collected papers after processing all steps of Kitchenham and Charter's methodology in column 'final number of collected papers' which is equal to 95.

## III. RELATED WORK

This section presents the related work on classifying and modelling of architectural attacks (ArchA). First, we present the existing classifications followed by the existing attack models.

### A. Classifications of Architectural Attacks

Until now, five classifications have been proposed for architectural attacks. Each proposal focuses on a different characteristic to organize the attacks. Figure 2 shows each classification and their metrics. Next, we describe each and analyse its benefits and limitations.

*1) Leakage-based Classification:* Based on the leakage source, the authors in [8] classified the architectural attacks of caches into three types: i) timing, ii) access, and iii) trace. In timing-driven attacks, the attacker performs an attack based on different temporal responses of components or operations. In access-driven attacks, the attacker performs an attack based on user behaviour of components or functions. Finally, in trace-driven attacks the attacker performs an attack based on high-level monitoring information from internal system monitors or some specialized external instruments (e.g., probing the power consumption).

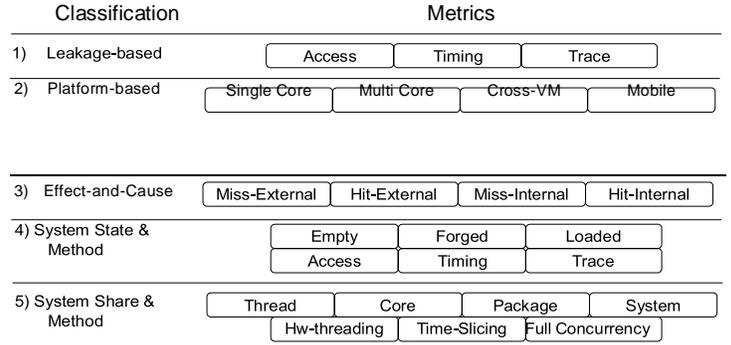

Fig. 2: Classifications of Architectural Attacks with Metrics.

However, novel attack techniques have emerged that do not exploit side-channel attacks. Hence, the above classification is not complete as it considers only information leakage. For example, Rowhammer attacks only inject faults in the main memory to gain privilege or alter system functionality. Consequently, this case has no leakage behaviour.

*2) Platform-based Classification:* In this classification scheme, the attacks are classified based on the hardware platform [12]. According to the authors, there are four possible platforms: i) single-core, ii) multi-core, iii) cross-VM, and iv) embedded/mobile. The single-core represents the most popular platform used in desktop/laptop computers. The attacks in this category require malicious processes to run in parallel with the victim processes as there is only a single core in the system. Such a malicious process is also defined as a spy process. The multi-core class includes desktop/laptop computers and more complex devices like servers. In this category, it is possible to attack using a spy process that runs on a core different from the victim or by forcing the attacker's application to run in a separate core. The third type, i.e. cross-VM, refers to virtualized systems. In a virtualized environment, the attacker and victim processes run in different virtual machines (VM). Hence, the attack has to interact with the other VM indirectly, which defines the term cross-VM. Even if the actual hardware is not disclosed in these systems, it is still possible to attack them by exploiting the vulnerabilities of VM implementations. The last category refers to embedded or mobile computer systems. These systems typically have limited performance as they are constrained devices. In this scenario, the operating system can be lightweight, or even the absence of one is a possibility. At the same time, the software typically has less restrictions to access the hardware configuration and low-level information.

Although the platform-based classification is interesting, it clarifies that several existing attacks might fit into two or more categories when the attacker can apply a specific attack. For example, the pure timing attack of Bernstein can be mounted on a single-core, multi-core, and embedded/mobile platforms. Therefore, the classification is not helping in understanding the basic principles of the attacks.

*3) Effect-and-Cause Classification:* The Effect-and-Cause classification proposed by [9] is dedicated to cache attacks. The effect refers to cache accesses, which can be a cache misses or a cache hits. The cause specifies which process made the effect happen; this is external when performed by the attacker process and internal when performed by the victim process. Consequently, this classification has four categories:
i) Miss-External, ii) Hit-External, iii) Miss-Internal, and iv) Hit-Internal. In the Miss-External class, attacks exploit cache misses caused by the victim process that are indirectly induced by the attacker. For example, in the Prime+Probe attack, the attacker accesses the cache before and after the victim's computation. Hence, the cache misses reveal the locations used by the victim. Note that the cache misses only occurred as the attacker already accessed these locations. Conversely, Hit-External exploits cache hits when the attacker accesses the cache. Similarly, the same idea is applied for Miss-Internal and Hit-Internal with the difference that the victim triggers the miss or hit behaviour. Pure timing attacks typically fall in the internal classes. The attacker only collects the final time of the operations, i.e., the variation in timing is a direct consequence of misses or hits during the victim's access.

This classification suits cache attacks but cannot be directly applied to general architectural attacks such as attacks related to processing or communication. Note that this classification can be updated by encompassing different architectural attacks by changing the options related to the effect.

*4) System State & Method Classification:* The System State & Method classification is proposed in [10] and uses two metrics to classify attacks. The first one is the system state before the victim's operation. The system state refers to the cache's content, which can be empty, forged or loaded. Empty state means that the cache is not initialised, and the initial state does not matter for the attack. The forged state refers to a cache initialised by the attacker, where the attacker writes specific contents to manipulate the behaviour of the cache. The loaded state refers to an initialized cache with some (or all) contents of the victim already loaded. As a result, the attacker can quickly identify changes in the cache. The second metric specifies the method used to collect the cache leakage. In this case, the same metrics of the Leakage-based classification are used, namely timing, access and trace.

Also this classification is specifically addressing cache attacks. As discussed in the introduction, ArchA can target any system component, including processing elements and communication structures. Hence, this classification is also limited for our purposes.

*5) Process Sharing & Method Classification:* The survey in [13] proposed a classification based on resource sharing and degree of concurrency. Sharing resources could occur on different levels such as thread, core, package or system level. Consequently, the classes related to sharing are: thread-shared, core-shared, package-shared, non-uniform memory ac-cess (NUMA)-shared and System-shared. With respect to the degree of concurrency, three categories are applicable; they are full concurrency (multi-core), time-sliced execution on a single core, and hardware threading (SMT). Note that this classification focuses on the relation between different processes running on a complex system and how vulnerabilities arise from such concurrent and sharing behaviour. In [13], only cache attacks have been classified under this scheme. However, this classification could embed many other logical side-channel attacks. Many attacks use a spy process (malicious process sharing the same core) or spy node (malicious core sharing the system). Although this classification can organize most of the logical SCAs, it does not help understand the main aspects of the attacks. It only defines the system configuration where attacks take place.

| Attack models | Phases | | | | |
|---|---|---|---|---|---|
| 1) Deng | Initial | Act | Observe | | |
| 2) Yarom | Monitor | Wait | Measure | | |
| 3) Osvik | Prime | Trigger | Probe | | |
| 4) Gruss | Access | Flush | Flush | | |
| 5) Kocher | Mount | Trick | Retrieve | | |
| 6) Bernstein | Learn | Attack | | | |
| 7) Bonneau | Manipulate | Observe | | | |
| 8) Canella | Prepare | Execute | Encode | Retrieve | Reconstruct |

Fig. 3: State-of-the-art Attack Models and their Phases.

*B. Attack Models*

In this section, we explain the state of the art ArchA models. They all divide the attacks into phases, i.e., a series of steps that describe an attack. Understanding and comparing the models is essential since different authors have different views of the same attacks; this results in models with different amount of phases. Even when authors agree on the number of phases, the phase definitions can be different. Figure 3 shows the attack models. As can be seen from the figure, the models use a different number of phases or different definitions of phases. Next, we describe the eight attack models.

*1) Deng:* The approach in [14] is related to side channel attacks on caches. In this work several well known attacks such as Flush+Reload, Evict+Time, Prime+Probe, Flush+Flush, Evict+Reload, Bernstein and Cache Collisions are analyzed. The approach divides these attacks into three steps: i) during the first phase, a memory access sets one single cache block in some known initial state; ii) during the second phase some action like fetch can be done by the victim or attacker;
iii) during the third phase a final action is taken to derive information by observing timing. Short time if there is a hit and longer time if there is a miss.

*2) Yarom:* The work in [15] analyzed the Flush+Reload attack on caches and is composed of three phases: i) during the first phase, the monitored memory line is flushed from the cache hierarchy; ii) the attacker then waits to allow the victim time to access the memory line before the third phase; iii) in the third phase, the attacker reloads the memory line, measuring the time to load it. If during the wait phase the victim accesses the memory line, the line will be available in the cache and the reload operation will take a short time. If, on the other hand, the victim has not accessed the memory line, the line will need to be brought from memory and the reload will take significantly longer.

*3) Osvik:* The approach in [16] divides Prime+Probe attacks into three steps: i) in Prime the attacker fills (parts of) the cache with data; ii) attacker triggers a sensitive operation (e.g. encryption); and iii) in Probe the attacker reads the data written in the Prime step and evaluates which addresses were used by the victim based on the observation of cache misses.

*4) Gruss:* The study in [20] analysed the Flush+Flush attack in caches with hierarchy. This approach decomposed the attack to three phases: i) at the first step, the attacker accesses the memory location, that is cached; ii) at the second phase, the victim only flushes the shared line. As the line is present in the last-level cache by inclusiveness, it is flushed from this level; iii) a bit also indicates that the line is present in the L1 cache, and thus must also be flushed from this level. To transmit a 0, the attacker stays idle. The victim flushes the line (step 1). As the line is not present in the last-level cache, it means that it is also not present in the lower levels, which results in a faster execution of the clflush instruction. Thus only the attacker process performs memory accesses, while the receiver only flushes cache lines. To send acknowledgment bytes the victim performs memory accesses and the attacker runs a Flush+Flush attack.

*5) Kocher:* Spectre, an attack which can be performed on CPU as well as caches is divided into three steps in [17] as follows: i) mounting, where the attacker introduces a sequence of instructions in the process address space; ii) trick, where the attacker induces the CPU to perform a transient execution; and iii) retrieve, where the attacker gathers the information through the covert channel.

*6) Bernstein:* Although, Bernstein attack is divided in 3 phases by the approach in [14], it is composed of two parts in [21]: i) a learning phase, where statistical reference models of the cache behavior are developed; and ii) an attack phase, where multiple encryption are collected and correlated with the reference models.

*7) Bonneau:* In [22] the Specter attack in AES is divided to two phases: i) Manipulation, in which the attacker manipulates the input message of the encryption algorithm to force collisions of cache addresses (cache hits) when the key hypothesis is correct; ii) Observation, in which the attacker observes if there is a reduction in the required traces to retrieve the key.

*8) Canella:* Work in [18] on the other hand divides Spec- tre and Meltdown to five phase: i) preparation of micro architecture; ii) execution for a trigger instruction; iii) transient instructions encode unauthorized data through a micro-architectural covert channel; iv) CPU retires trigger instruction and flushes transient instructions; v) reconstruct secret from micro architectural state.

As discussed there is no general formal structure for modeling the attack phases. In some articles, authors split particular attacks into two phases, while in others they split them into three phases. A unified attack model that covers all the ArchA is currently missing. This implies that the underlying patterns that reveal the attack mechanisms are not well-defined and understood.

To exceed the state of the art, this work proposes first a new taxonomy for ArchA using three metrics: what, where and how. These metrics can cover all the architectural attacks. Secondly, this work propounds an innovative attack model which contains five steps: Setup, Trigger, Operation, Retrieve and Evaluation. In the following sections these aspects are further presented in more detail.

## IV. PROPOSED ATTACK MODEL

This section introduces our proposed attack model, a formalized and generic structure to describe Architectural Attacks behavioural patterns. Our model organizes any attack as a five step process. Equation 1 defines our *Attack Formula*, as a set {S, T, O, R, E} where S, T, O, R and E stands for Setup, Trigger, Operation, Retrieve and Evaluation phases, respectively.

$$AttackFormula = \{ S \quad T \quad O \quad R \quad E \} \quad (1)$$

Next, each phase is detailed:

1) *Setup*: The first phase is the preparation of an attack. In other words, to put the system in a known state suitable for the attack.
2) *Trigger*: The second phase regards the action(s) to activate the target vulnerability.
3) *Operation*: The third phase relies on the victim. In this phase, the victim executes and expose the target vulnerability. Such vulnerability can be a direct (i.e., perform some operation) or indirect (i.e., side-channel leakages during operation) action. Note that this is the only phase in which the victim operates.
4) *Retrieve*: The fourth phase gathers the necessary information from the system to evaluate the success of the attack. Examples of collected information are operation time, cache set status or privilege bits values.
5) *Evaluation*: The final phase analyzes the collected data from *Retrieve* phase to obtain the final attack objec- tive, which can be IP design information, Functionality changes, or personal Data.

By using our attack model, attacks can be structured in a generic and formal representation. Consequently, there are two main benefits. First, the main actions to perform the attack are highlighted and separated according to their primary objective (i.e., preparing the system (Setup), triggering the attack (Trigger), exposing the vulnerability (Operation), or retrieving the information (Retrieve)). Taking the phase into consideration, a designer could elaborate a strategy to prevent attacks. For example, the system could avoid the Setup actions targeted by the attacker. Another example would be to find a way to prevent the vulnerability related information can be retrieved by the attacker. Therefore, engineers can investigate the best alternative to avoid attacks (all or a subset).

The second benefit of our attack model regards the generic representation, which allows comparing different attacks. Such comparisons make it possible to identify common vulnerabilities exploited by different attacks. Hence, a designer can elaborate more efficient countermeasures.

## V. Proposed Taxonomy

Architectural Attacks are software attacks that exploit any hardware vulnerability. Modern integrated circuits like Systems-on-Chip or Multi-Processors Systems-on-Chip are defined as complex components, from hardware accelerators to memories and interfaces. Hence, architectural attacks comprise a wide range of possibilities as an adversary might theoretically exploit any hardware components.

In the literature, there are several software attacks targeting hardware components. However, due to a very different nature among hardware components, like memories, processors and accelerators, the attacks were organized in separate classifications. In the previous section, we observed that memory attacks are classified in the cache attack category, processor attacks in the transient execution category, while others might fall into more generic concepts like Side-Channel Attacks. In order to bring all architectural attacks to the same scheme, this paper proposes a novel *taxonomy*. As a result, our metrics and classification provide a systematic way to analyze and evaluate such attacks.

Our proposed taxonomy uses three different metrics as criteria:
- Target: **What** the attacker is looking for.
- Location: **Where** is the victim's vulnerability.
- Method: **How** the attacker exploits the vulnerability.

Figure 4 presents the taxonomy as a hierarchical arrangement of our proposed metric criteria. Additionally, we show where all attacks evaluated in this survey are located in the taxonomy. In the following subsections, each criteria is detailed.

### A. Target (What)

The attack Goal is the main objective behind the attack. Three options comprise all alternatives, and they are:
- IP: The attack looks for intellectual property information, such as implementation details of a design and some specific software/application. Engineering information is valuable in the global market, which avoids months of research and development and reduces risks when building something new.
- Functionality: In this case, an attacker targets to modify temporarily or permanently the functionality of a system. Some examples of functionality exploitation are bypass a security check, provide privilege escalation, decrease overall system reliability.
- Data: When the goal is data, the attacker aims to steal or corrupt data inside a system, which in most cases, important information are encrypted. Hence, most data attacks target first to retrieve the secret key of the system.

### B. Location (Where)

The location refers to where the vulnerability is. Since architectural attacks are software exploiting hardware, location criteria regard only to hardware components. Since the different type of components exists in hardware, we classify these criteria into three parts:
- Processing: Processing elements are processors, hardware accelerators and co-processors. Any element responsible for an operation or task in the system can be classified as Processing.
- Memory: Memory relies on storage components. Some examples are cache memories, flash memories, read-only memories (ROMs), static and dynamic RAMs (SRAMs and DRAMs).
- Communication: Communication components are the ones responsible for interface other components. Internally, the communication is provided by a bus system (e.g., AMBA AHB or AMBA AXI) or even a network-on-chip (NoC). These components interface the internal elements - Processing and Memory - among each other. Moreover, there are communication components dedicated to external interfaces. Examples are Ethernet, UART serial, and Bluetooth.

### C. Method (How)

These criteria focus on how the attacker triggers a specific vulnerability present in the hardware. Three options are given:
- Injection: In this method, an attacker can force a trigger to exploit a certain vulnerability. Fault attacks like Buffer Overflow use the injection method, where the adversarial overwrites part of the memory to change the victim's behaviour.
- Manipulation: In this method, an attacker manipulates a component from the system to trigger the vulnerability. This method is employed when the attacker cannot create or force a trigger to exploit a certain vulnerability. There are several reasons for that, but the main reason relies on permission accesses.
- Observation: The observation method occurs when the vulnerability does not need a trigger. In this case, the vulnerability is always present, i.e., the victim is always leaking information. In this method, the attacker only needs to find a mean to monitor the victim's behaviour and process the leakage into meaningful data.

## VI. IP attacks

IP attacks target extracting sensitive information related to an engineering process. Such attacks aim to gain a technical advantage whether to replicate, fake or reuse a design. Examples of sensitive technical information are the algorithm used, the division between software and hardware for a certain task, or even non-disclosed functions inside the system. Therefore, this section focuses on software attacks that can extract sensitive technical data, also known as Intellectual Property. Figure 4 presents the references of existing IP ArchA under the 'Intellectual Property (IP) attack' branch.

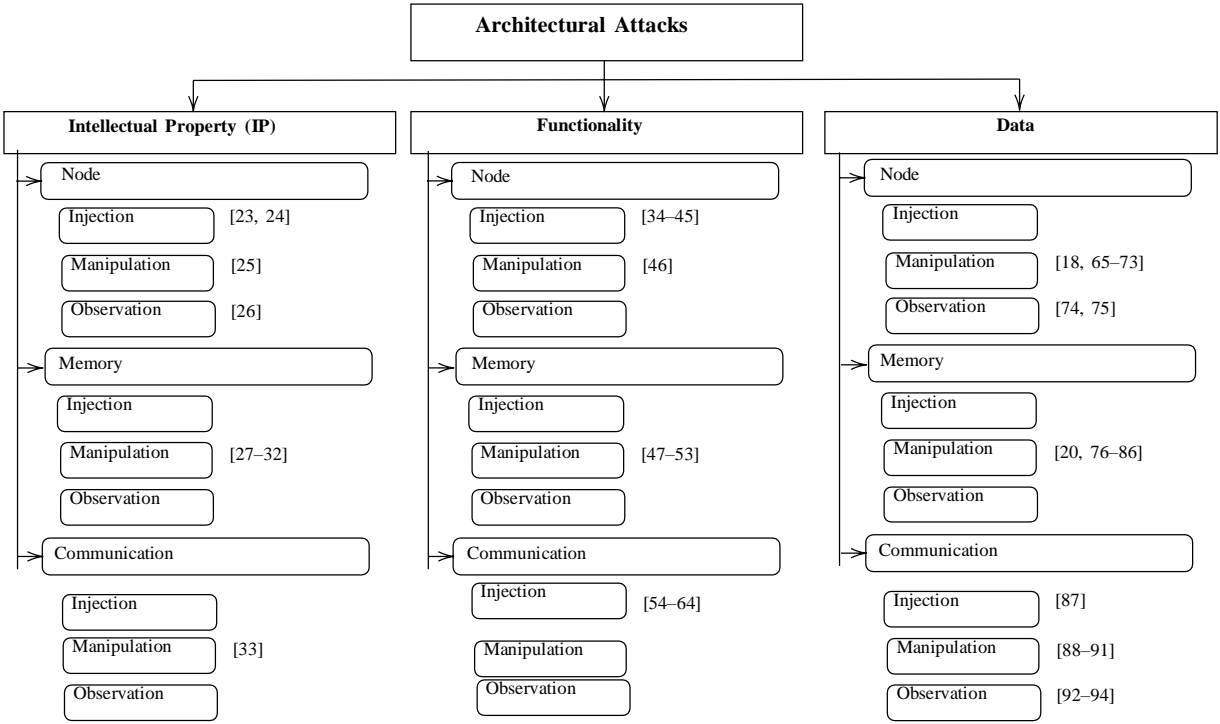

Fig. 4: Proposed Architectural Attacks Taxonomy

*A. Node:*

This subsection presents the attacks that aim to reveal design secrets from the nodes, like processors and graphical processing units (GPUs). According to our taxonomy, we categorize node attacks into three main groups, namely injection (suffix -INJ), manipulation (suffix -MAN) and observation (suffix -OBS). Next, each group is detailed.

**Group IP-NODE-INJ.**

This group contains attacks that generally injects random inputs to force the processor into abnormal conditions, revealing unexpected features or instructions. Later on, the attacker monitors the processor behavior to discover anomalies.

**Attack Formula:** Steps of the proposed attack formula are described as follows:

- S: In the setup phase, the attacker defines the main parameters to be explored in a node architecture. Examples are instruction opcodes, microcode values, special hardware parameters.
- T: The attacker crafts new data based on the defined parameter to the attack and applies in the node architecture. This crafted information is out of the specification, which configures an injection technique.
- O: The node executes or tries to execute the input provided.
- R: The output behaviour of the node is observed, which can be an expected output value or just different timing behaviour.
- E: In this phase, when the system does not crash and present a different behaviour, the attacker knows a valid input was revealed. Next, he/she analyzes what function has been uncovered by exploring the same parameter under different circumstances.

**Attacks:** In [23], the author presents a tool called Sandsifter. It audits x86 processors for hidden instructions and hardware bugs by systematically generating machine code to search through a processor's instruction set and monitoring execution for anomalies. This attack injects random inputs to force the processor into abnormal conditions. In [24], the attacker creates microcode updates to two AMD processors (K8 and K10) and observes the output. The output is represented by the register values and memory locations. The underlying idea is to generate distinct behaviours between the original and the patched macroinstruction execution. More precisely, the patch contains a microcode instruction that constantly crashes on execution.

**Group IP-NODE-MAN.** This group refers to attacks that force specific behaviours to infer design-related information about the node. Differently from the IP-NODE-INJ group, this group creates only expected and valid inputs. Hence, it falls into the manipulation category. The objective here is to define

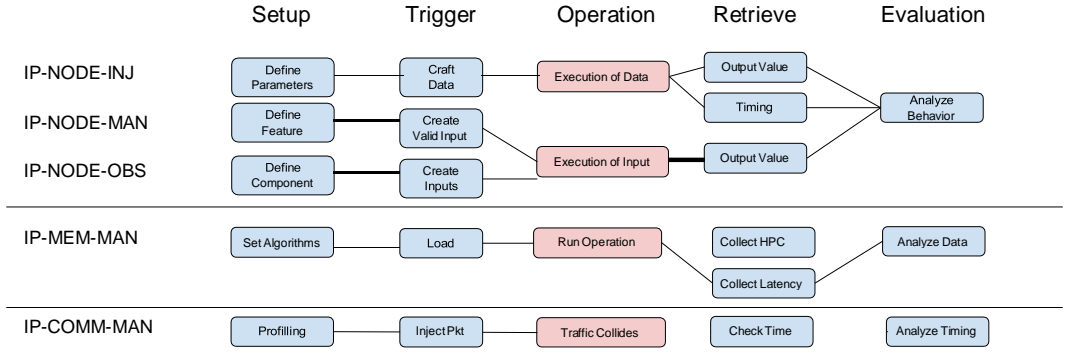

Fig. 5: Attack Phases of IP Attacks and the associated groups.

a feature to observe and provide the input that maximizes the actuation of it.

**Attack Formula:** Steps of the proposed attack formula are described as follows:

S: In the setup phase, the attacker defines which feature to highlight during the attack. Examples are schedule algorithms, privilege definitions, or arbitration of tasks.

T: The attacker creates valid inputs that can highlight the target feature and apply it to the node.

O: The node executes the provided input.

R: The execution of the task will reveal the strategy or algorithm behind the design. Multiple attempts can be made to identify these differences.

E: In this phase, the attacker analyzes the outputs to infer what known algorithm matches with the observed result. Finally, the attacker understands important design-related information.

**Attack:** Last attack proposed in [25] runs specific bench-marks into the GPU to infer design-related information. Each benchmark aims to emphasize a specific characteristic of the GPU, hence putting in evidence how it works. Since this attack runs expected programs but pick carefully the ones that might reveal design secrets, we define this attack into the manipulation category.

**Group IP-NODE-OBS:** This group contains attacks that mainly observe the system output when regular input is provided. Unlike the group IP-NODE-MAN, this group does not create particular inputs to manipulate the system to reveal more information. The attacks in this group focus on running typical applications or just accessing the system components and observe the behaviour, output values, or timing.

**Attack Formula:** Steps of the proposed attack formula are described as follows: model as follows:

S: In the setup phase, the attacker defines which component to explore.

T: After the setup, the attacker creates many different inputs and apply to the system. In this step, the attacker can use random input generation, just trial and error, or brute-force all possibilities.

O: The node executes the provided input.

R: The execution of the task will reveal the strategy or algorithm behind the design. Multiple attempts can be made to identify these differences.

E: In this phase, the attacker analyzes the outputs to infer what known algorithm matches with the observed result. Finally, the attacker understands important design-related information.

**Attack:** In [26], the attacker aims to discover which special registers can trigger hidden x86 instructions. The registers able to perform such an operation are the global configuration registers, known as MSR registers in the x86 architecture. The first step in this attack performs a timing analysis accessing all available MSR register to understand which one contains a unique behaviour. As most MSR registers trigger similar functions in the system, their access time is also similar. Only the different MSRs will cause a different access time.

*B. Memory:*

Existing attacks to memory design information fall only on the manipulation category, defined here as group IP-MEM-MAN.

**Group IP-MEM-MAN:** The attacks in this group create specific programs to run in the system, and after that, collect running information to infer memory characteristics. Attacks in this group can target cache memories and DRAMs. Most attacks targets on features like physical-to-logical address mapping, cache line size and replacement policies, as well as physical characteristics like access latency.

**Attack Formula:** Steps of the proposed attack formula are described as follows:

S: Prepare set of algorithms to highlight the target characteristic.

T: Run the algorithm in the system.

O: The natural execution of the algorithm create covert channels, like different latency's or different behavior(that can be observable).

R: Retrieve: Attacker access system monitors like High Performance Counters or use its own measurement means (e.g., timer) to collect side-channel information.

E: The attacker process the obtained information to infer the feature value. The attack is repeated to guarantee that the findings are highly provable. Fine tuning in the set of algorithms to optimize the new runs are also possible.

**Attacks:** There are several research efforts that infer properties of caches using measurement-based analysis [27, 28, 31]. These solutions make use of performance counters available in current platforms to infer properties of the cache hierarchy. Recent works that infer DRAM properties [29, 30, 32] focus on the address mapping, like virtual-to-physical memory allocation scheme by first inferring the mapping between virtual address bits and physical bank bits for the Intel Xeon processor using latency-based analysis. Another example uses a latency based analysis to identify channel, rank, and bank bit mapping between virtual and physical addresses.

*C. Communication:*

Only manipulation attacks have been proposed to perform reverse engineering on the communication structure to the present moment. We describe this group IP-COMM-MAN in the following.

**Group IP-COMM-MAN:** This group refers to attacks that inject multiple different packets to infer the configuration used inside the communication.

**Attack Formula:** Steps of the proposed attack formula are described as follows:

- S: The attacker requests the packets in this step.
- T: By responding to the request and sending the packets to the network, this step is completed.
- O: In this step, the flooding is performed. This can be done by requesting too many packets, triggering packets with wrong paths, triggering packets that intent to create a deadlock, triggering packets that can not reach the target and etc.
- R: Flooding causes the victims to lack resources.
- E: The availability of the victim's resources is checked in this step. If there is a lack, this means the DoS attack has performed successfully.

**Attack:** The attack presented in [33] showed a algorithm to uncover the details of the communication structure by simply manipulating the packets.

Figure 5, demonstrates different steps of IP attacks. As can be seen in this figure some attacks have steps in common. For instance 'IP-NODE-MAN' and 'IP-NODE-OBS' have similar Operation, Retrieve and Evaluation steps. Obviously, this kind of analysis and revealing the detailed information on the structure of attacks can lead to building a better mitigation strategy. In the next steps the same analysis is reported for Functionally and Data attacks.

## VII. FUNCTIONALITY ATTACKS

Depending on the target component, the attack behavior will differ, and therefore, we identified these behavior in the following. Figure 4, represents the existing attacks regarding this attack type under the 'Functionality attack' branch.

*A. Node*

Functionality attacks to processing elements aim to subvert the program's control flow from its normal course, and force such program to act in the manner of attackers will. In the attacks observed in the state-of-the-art that focus processing elements, we identified three different attack patterns, namely buffer overflows, reuse of existing code and speculative overflows. Each pattern is described in the following.

**Group FUNC-NODE-INJ:** The first group refers to the attacks known as buffer overflow. Buffers are memory storage regions that temporarily hold data while it is being transferred from one location to another. A buffer overflow (or buffer overrun) occurs when the volume of data exceeds the storage capacity of the memory buffer. As a result, the program attempting to write the data to the buffer overwrites adjacent memory locations. Attacks which exploit memory errors such as buffer overflows constitute the largest class of attacks reported by organizations such as the CERT Coordinatione Center [95], and pose a serious threat to the computing infrastructure. In the literature, various kind of Buffer overflow attacks [39] and [44] are described as follows:

i) Buffer overflow which overwrites the return address,
ii) Buffer overflow which overwrites the frame pointer,
iii) Buffer overflow which overwrites the function pointer,
iv) Buffer overflow which overwrites the dynamic Linker Tables.

However, we believe that the above attacks can be identified in one main group since their attack Algorithms are almost the same and they differ only in Trigger phase.

**Attack Formula:** The attack phases are as follows:

- S: Abuse input functions w.r.t type size, for instance if the size of the buffer is set to 500 characters, the attacker insert 510 characters in order to overrun the buffer (stack).
- T: After the setup, the attacker will overwrite the sensitive area of code and point it to an exploit payload, to gain control over the program. This can cause the program to behave unpredictably and generate incorrect results, memory access errors, or crashes. Note that, this phase is the only phase where the mentioned attacks slightly differ.
- O: This part redirect execution to another attacker's code giving privileges. For example, an attacker can overwrite a pointer (an object that points to another area in memory) and point it to an exploit payload, to gain control over the program.
- R: Attacker's code execute
- E: Checking privileges

**Attacks:** Works in [39] and [44], exhaustively studied the different kind of buffer overflow attacks. Study in [34] demonstrates security attacks that exploits unsafe functions to overflow stack buffers and methods to detect and handle such attacks. Approach in [45] argues that the defense such as Control-Flow Integrity (CFI) does not stop control-flow hijacking attacks which occurs by Buffer overflows (return address overwritten). It demonstrates a mitigation strategy for such a case. Study in [35] exploits different protection techniques for stack based buffer overflows and consequently presents four tricks to bypass these protections. Authors in [36] present address obfuscation strategy to prevent these kind of attacks. Work in [37] studies different buffer overflow attack and a mitigation method at compile time for address randomization to stop the attack. The work in [38] demonstrates a code injection attack which is a result of a buffer overflow and a mitigation strategy by Instruction Set Randomization (ISR). [40] introduced a mitigation technique called SmashGuard which is a hardware solution to prevent manipulation of buffer overflow attack on function return address. Study in [41] is a solution to detect code reuse attacks on ARM mobile devices. Work in [42] describes buffer overflow attacks(Stack Smashing) in the UNIX operating system. And finally, authors in [43] introduced another mitigation techniques called StackGuard to prevention of buffer-overflow attacks.

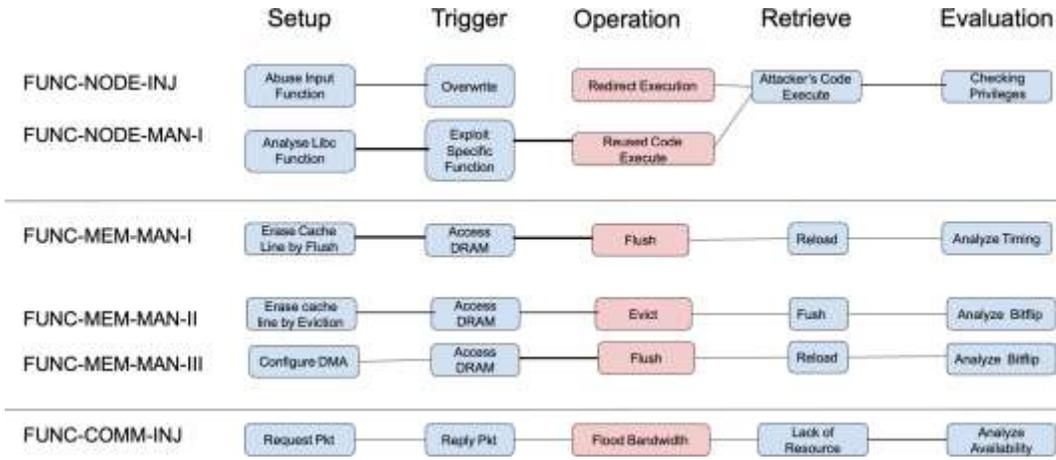

Fig. 6: Attack Phases of Functionality Attacks and the groups.

**Group FUNC-NODE-MAN:** This group refers to attacks that reuse existing code, also known as Return-oriented Programming attacks or Return-into-libc attacks. This type of attack is used when a system mitigates the possibility to inject code into memory to be executed. The most used protection in operating systems that prevents code injection is the "W⊕X" [96], which prevents that writable data in memory to be executable at the same time. As a response, attackers reuse existing code to perform valid actions into the system. The standard C library, libc, is the most common target as it is placed near the kernel code and provides very useful functions like system calls. However, in principle, any available code, either from the program's text segment or from a library it links to, could be used. Consequently, the attacker can identify in the code useful sequences that when put together create the malicious functionality. Each sequence can perform a specific function like addition of two registers, load constant, load memory data, store, etc. Note that it is important that each code sequence ends with a return call. Then, the attacker only manipulates the stack to call each sequence (i.e., jump to the address of the first instruction of the sequence) in a specific order to achieve the desired functionality. The stack manipulation can use buffer overflows attacks previously mention as Group A.

**Attack formula:** In this group, the steps of formula are as follows:

S: In the setup phase, the attacker dumps the memory content and identifies useful code sequences. Thereafter, the attacker builds a "program" ordering the sequence addresses he must call.

T: Attackers manipulates a function call, typically performed through buffer overflow attacks.

O: This part represents the processor executing the instructions of each sequence (i.e., valid code already in the system).

R: There are many exploitation methods for this attacks since the attacker defines the algorithm to be executed, but we will consider in this example a privilege escalation of the attacker's user.

E: Verify the level of privilege in the system.

**Attacks:** Work in [46] describes this kind of attack and a mitigation procedure to prevent control-flow hijacking attacks and code-injection attacks. The work introduces KAISER as a algorithm to strict kernel and user space isolation such that the hardware does not hold any information about kernel addresses while running in user mode.

### B. Memory Element

The main series of functional attack on memory typically comes from a same family of attacks called Rowhammer. Rowhammer is a security exploit that takes advantage of an unintended and undesirable side effect in dynamic random-access memory (DRAM) in which memory cells interact electrically between themselves by leaking their charges, possibly changing the contents of nearby memory rows that were not addressed in the original memory access. These attacks perform repeatedly access to certain memory rows with a high frequency in order to degrade the internal charging capacitance. Such operation is defined as "hammering" and that is why such threat is called as Rowhammer. There are three possible attack patterns, the flush based [47], eviction based [50] and remote based [52]. They are described in details in the following.

**Group FUNC-MEM-MAN-I:** The first group perform "hammering" through a flush operation. Flush operations can erase a cache line or region. By continuously forcing flush operations, the same memory rows have to be accessed to retrieve the missing information.

**Attack Formula:** Steps of the proposed attack formula are described as follows:

- S: In the setup phase, the attacker access to a memory row and erase cache line through a Flush instruction.
- T: After the setup, attacker access to the next memory row and flush this line as well.
- O: This phase represents a Flush. It is applied to determine whether the data from the previous step has changed or not. This phase helps the Evaluation phase to gather the information related to identified the cache timimg.
- R: This phase is performed by Reload instruction.
- E: In this phase the timing behavior is analyzed. If during the Trigger phase, the victim accesses the memory line and flush it, the line will need to be brought from memory and the reload operation will take a longer time. If, on the other hand, the victim has not flushed the memory line, the line will be available in the cache and the reload will take significantly shorter time.

**Attacks:** The known attacks belong to this groups are as follows. Work in [47] introduces a flush based hammering technique called SGX Bomb, which is able to lock down the processor. DRAMMER the hammering attack in [48] relies on the predictable memory reuse patterns of standard physical memory allocators which is implemented on Android/ARM. [49] presents DRAMA attacks, a class of attacks that exploit the DRAM row buffer that is shared in multi-processor systems

**Group FUNC-MEM-MAN-II:** The second group perform "hammering" by causing evictions in the cache memory. This can be done by accessing data that aligns to the same line in the cache. As a result, the victim address in the DRAM have to be continously accessed due to the eviction in the cache. In comparison to the flush based attack, this approach is much less efficient as it requires multiple access to DRAM to successfully evict a cache line.

**Attack Formula:** Steps of the proposed attack formula are described as follows:

- S: In the setup phase, the attacker access to a memory row and erase cache line through an Eviction instruction.
- T: After the setup, attacker access to the next memory row and evict this line as well.
- O: This part represents Eviction instruction. They are applied to determine whether the data from the previous step has changed or not. This phase helps the Evaluation phase to gather the information related to identified the bitflip.
- R: This phase is performed by Flush.
- E: This phase checks if there is a bit flip after all the above four steps.

**Attacks:** Study in [50] is an eviction based hammering attack which causes the bit flip by hammering only one location in memory. [51] demonstrated that an attack called Rowhammer.js can be forced into fast cache eviction to trigger the Rowhammer bug with only regular memory accesses.

**Group FUNC-MEM-MAN-III:** The third group is performed remotely. Note that in some works this concept is extended to Remote Direct Memory Access (RDMA). In other terms, it is applied on a server i.e., it continuously receives packets from sender and write on RDMA. In this group for the Operation phase is done by Flush, Trigger phase by a Relaod instruction and Evaluation phase by analyzing cache timing behavior.

**Attack Formula:** Steps of the proposed attack formula are described as follows:

- S: In the setup phase, the attacker access to a memory row. It can be done by read or write instruction.
- T: After the setup, attacker access to the next memory row.
- O: In this part a Flush instruction is executed. This phase is done by the victim lead the attacker to gain access to some information.
- R: A Reload is applied to determine whether the data from the previous step has changed or not. This phase helpsthe Evaluation phase to gather the information related to measuring the time to load it.
- E: In this phase the timing behavior is analyzed.

**Attacks:** Throwhammer [52] and Nethammer [53] are two examples of this group of attacks. These attacks are able to remotely perform hammering on servers by continuously receives packets from senders and write on RDMA..

*C. Communication*

This subsection focus on functionality attacks in the communication structure. Observing the existing attacks, only one group could be defined, which falls into injection category.

**Group FUNC-COMM-INJ:** Denial of Service (DoS) attacks are exhaustively used in the state of the art when the vulnerability of a communication protocol is assessed. This group of attacks generally disrupts the system by overloading resources such that a malicious application generating packets with a high injection rate to produce this attack. However, in some cases the attack is able to overload the communication infrastructure.

**Attack Formula:** Steps of the proposed attack formula are described as follows:

This attack generally attempts to flood the system by sending massive amount of packets. It is notable that different flooding algorithms have been demonstrated by researches, for instance in some works, the flooding is performed by assigning a wrong path to the packet. This means, a packet with erroneous paths introduced to the network with the aim to trap it into a dead end. In other works, packets with paths that intentionally disrespect deadlock-free rules of the routing technique are used. These packets intent to create deadlocks in the network. While, in some studies, packets can't reach their targets and stay turning infinitely in the network, causing a waste of bandwidth, latency and power [58]. Although the aforementioned algorithms are different in flooding the network we believe, they can be divided into similar phases. In the following the proposed five phases for this group are describe.

- S: This step is performed by the attacker and request for the packets.
- T: This step is done by replying to the request and send the packets to the network.
- O: In this step the flooding is performed. This can be done by requesting too many packets, triggering packets with wrong paths, triggering packets that intent to create deadlock, triggering packets that can not reach the target and etc.
- R: Due to the flooding the lack of resources for the victim occurs.
- E: In this step availability of the resource for the victim is checked. If there is a lack, this means the DoS attack has performed successfully.

**Attacks:** The method in [54] showed that a thread running on an implementation of a SMT processor can suffer from DoS by a malicious thread. This work proposes a number of algorithms to counter such attack: some affect the core scheduling algorithm and others simply attempt to identify activity that would affect threads sharing the same processor core. The proposed attack in [55] performs packet inspection and injects faults to create a DoS attack. Faults injected are used to trigger a response from error correction code schemes and cause repeated retransmission to starve network resources and create deadlocks capable of rendering single application to full chip failures. [56] is a mitigation approach for DoS diagnosis scheme which detects DoS attacks based on the performance degradation of sensitive flows. The proposed method using latency metrics can leveraged for detecting a DoS attack, and also, aiming to locate the attack source. Study in [57] identifies a DoS attack and proposes to protect commu- nication and computation against this attack. This is done by creating continuous Secure Zones at runtime. These zones are isolated area in the system, preventing traffic flows to cross the boundaries of the zone and thus a secure application can be executed securely. Approach in [58] presents a monitoring sys- tem for NoC based architectures, whose goal is to detect DoS attacks carried out against the system's information. Method in [59] is another DoS mitigation which creates security zones according to the security requirements of the applications. Proposed work in [61] can detect DoS attack on NoC based on evaluating runtime latency. This evaluation enables monitoring the trustworthiness of the NoC throughout the chip lifetime. Authors in [60] introduce DoS attacks on NoCs used in SoC design. These attacks have different implementations such as full ASIC implementation and full FPGA implementation, as well as their mitigation strategies. Proposed method in [62] presents a mitigation strategy based on the isolation of some secure zones that is in response to the current DoS strategies. Study in [63] proposes a hardware mechanism to secure the data transactions between the NoC routers. The security mechanism is capable of detecting and preventing DoS attack that aim on degrading the system performances. A firewall is demonstrated in [64] for a hardware-based NoC which performs rule-checking of memory requests at segment-level. This firewall aims at protecting NoC against DoS attacks on ARM and Spidergon STNoC.

Fig. 6 shows steps of Functionality attacks and demonstrate which types have similar structure.

## VIII. DATA ATTACKS

This section describes different attacks which mainly are performed on data. Figure 4 presents the references of existing Data ArchA under the 'Data' branch. The overall view of this section is demonstrated in Figure 7.

### A. Processing Element

In the attacks observed in the state of the art that focus pro- cessing elements, we identified seven different attack patterns. Each pattern is described according to our attack phase model and is depicted in Figure 7. In the following, each group is detailed.

**Group DATA-NODE-MAN-I:** The first group refer to attacks on the branch prediction unit (BPU) of modern processors. In summary, they exploit the speculative behaviour of such component to force illegal instructions to execute. Although the processor can detect such mistake caused by speculation and undo the illegal operation, the processor or the system might have changed. If carefully designed, the attacker can hide the stolen information in such different states of the system. The most typical example is to use the cache memory, where the access to an address will permanently change the cache state.

**Attack Formula:** In the following the phases of this attack has been presented:

- S: In the setup phase, the attacker performs many executions forcing the branches to take the same result always (e.g., taken). As a result, the BPU will be manipulated to speculate to such a condition.
- T: After the setup, the attacker will execute an illegal instruction since the BPU is conditioned to execute it speculatively.
- O: This part represents the processor executing the opera- tions of the illegal instruction, and before it finishes, undo all operations due to exception handling. Such executed instruction that was undone is also defined as transient execution.
- R: The illegal instruction performed an operation reflected in a change of specific cache address. This address was defined by one bit of sensitive data. The target bit can be used in a shift operation which may point to zero when the bit is zero, or address 64 when the bit is 1 (considering a shift by 6).

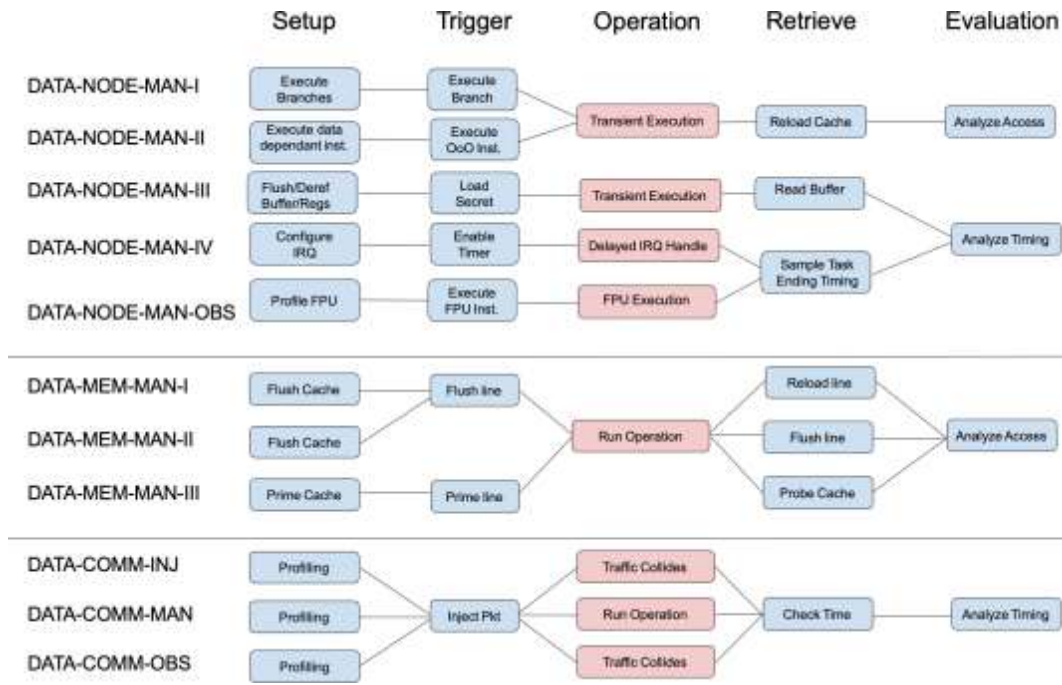

Fig. 7: Attack Phases of Data Attacks and the groups.

E: Identify if address zero or address 64 has a different state. The used address will present a hit behaviour.

**Attacks:** Approach in [73] presents Spectre attack which is one of the well known BPU attacks. This attack involves inducing a victim to speculatively perform operations that would not occur during correct program execution and which leak the victim's confidential information via a side channel to the adversary. Netspectre is described in [66] which follows the same idea of Spectre but over a network. Work in [71] is another type of these attack called BranchScope where the attacker infers the direction of an arbitrary conditional branch instruction in a victim program by manipulating the shared directional branch predictor.

**Group DATA-NODE-MAN-II:** The second group attack are the out-of-order execution. Any modern processor rearranges the order of instructions to be executed to add a gap between instructions with data dependency. Such filling instructions must be independent of the current flow. Consequently, those filling instructions are executed out-of-order, and in the last part of the pipeline, a module reorganizes the outputs (i.e., put again in order). However, a common vulnerability inside most processors is that illegal operations are just checked in the last part when the reordering happens. As a result, this causes a transient execution. With this understatement, these kind of attacks organize a program where the independent instruction performs an illegal operation with sensitive data. Such operation typically accesses the cache, which modifies its state. Hence, even with the undo performed by the processor, the cache state is already changed. Finally, the attacker needs to reread the cache and verify what address the state has changed.

**Attack Formula:** In the following the phases of this attack has been presented:

S: In the setup phase, the attacker runs several instructions with data dependency.

T: After the setup, the attacker will execute an illegal instruction independent of the previous instructions. The out-of-order unit will prioritize this independent instruction.

O: This part represents the processor executing the operations of the illegal instruction, and before it finishes, undo all operations due to exception handling. Such executed instruction that was undone is also defined as transient execution.

R: The illegal instruction performed an operation reflected in a change of specific cache address. This address was defined by one bit of sensitive data. The target bit can be used in a shift operation which may point to zero when the bit is zero, or address 64 when the bit is 1 (considering a shift by 6).

E: Identify if address zero or address 64 has a different state. The used address will present a hit behaviour.

**Attacks:** One of the well known attack belong to this group is Meltdown [69]. This is a hardware base attack which works on different Intel microarchitectures and exploits side effects of out-of-order execution on processors to read ar-bitrary kernel-memory locations including personal data and passwords. In [18] two new version of Meltdown have been proposed. Meltdown-PK (Protection Key Bypass) on Intel, and Meltdown-BND (Bounds Check Bypass) on Intel. Contrarily, approach in [70] presents Foreshadow, a software-only mi-croarchitectural attack that dismantles the security objectives of current SGX implementations.

**Group DATA-NODE-MAN-III:** This type of attacks targets the internal storage elements of the processor. Both registers and buffers can be exploited by these attacks, mainly related to load and store memory data. Some examples are the Line-Fill Buffers, Load Ports and Store Buffers. Store Buffers (SBs) are internal buffers used to track pending stores and in-flight data in optimizations such as store-to-load forwarding. Some modern processors enforce a strong memory ordering, where load and store instructions that refer to the same physical address cannot be executed out-of-order. However, as address translation is a slow process, the physical address might not be available yet. The processor performs memory disambiguation to predict whether load and store instructions refer to the same physical address. This process enables the processor to execute load and store instructions out-of-order speculatively. As a micro-optimization, if the load and store instructions are ambiguous, the processor can speculatively store-to-load forward the data from the store buffer to the load buffer. On the other hand, Line Fill Buffers (LFBs) are internal buffers that the CPU uses to keep track of outstanding memory requests and perform a number of optimizations such as merging multiple in-flight stores. Sometimes, data may already be available in the LFBs and, as a micro-optimization, the CPU can speculatively load this data (similar optimizations are also performed on e.g. store buffers). In both cases, modern CPUs that implement aggressive speculative execution may speculate without any awareness of the virtual or physical addresses involved [67].

**Attack Formula:** In the following the phases of this attack has been presented:

- Setup: In this step, victim as part of its normal execution, loads or stores some secret data. The secret data can be a piece of leaked data from inactive code by forcing cache evictions.
- Trigger: In Trigger, attacker also performs a load such that the processor speculatively uses data from the LFBs rather than valid data.
- Op: When the processor eventually detects the incorrect speculative load from the pervios step, it discards any and all modifications to registers or memory, and restart execution with the right value.
- Retrieve: Since traces of the speculatively executed load still exist at the microarchitectural level, we can observe the leaked data using a simple (Flush + Reload). Thus, the attacker reloads the data (memory line), together the information related to measuring the time to load it.
- Evaluation: In the Evaluation phase, the timing behavior is analyzed.

**Attacks:** RIDL is an attack of this type introduced in [67]. RIDL attacks are implemented from linear execution with no invalid page faults, eliminating the need for exception suppression mechanisms and enabling system-wide attacks from arbitrary unprivileged code (including JavaScript in the browser). Fallout is another attack presented in [68]. Fallout, is transient execution attack that leaks information from a previously unexplored store buffer. An unprivileged user processes can exploit Fallout to reconstruct privileged information recently written by the kernel.

**Group DATA-NODE-MAN-IV:** This group contains attacks that exploit interruption handling to monitor the time of sensitive operations indirectly. When a high priority interruption occurs, the processor finishes the current execution and changes its context to the interruption handling. However, depending on the operation that has been executed inside the processor, the interruption handling will take different execution times. These time differences result from a delayed process to change the context, which correlates with the target operation. Consequently, an attacker can sample all these timings and perform an analytical approach to infer a secret.

**Attack Formula:** In the following the phases of this attack has been presented:

S: Configure an interruption component to periodically activates.

T: Enable the component responsible for interrupting the processor. Typically, a timer is used in this step.

O: Processor receives the IRQ signal, finishes the current instruction execution, changes the processor context, and executes the IRQ handling function. Such steps may result in different timing due to the instruction and data being operated.

R: Save the time to finish the IRQ handling.

E: After several attempts, all saved time data can be analyzed to infer sensitive information. A computation based on correlation can be applied here.

**Attacks:** Only one attack has been published on this domain. The authors in [65] present Nemesis, an attack that abuses the CPU's interrupt mechanism to leak instruction timings from CPU executions. Basically, the approach uses a timer to create an interruption based attack.

**Group DATA-NODE-MAN-V:** This identified group targets co-processors or hardware accelerators. Commonly, specialized hardware components are designed to obtain maximum performance, and therefore, present commonly different timing behaviour depending on the input data. Regarding processing elements, the most common attack refers to Floating Point Units (FPUs).

**Attack Formula:** In the following the phases of this attack has been presented:

S: Profile the hardware component. In the case of FPU, random inputs are applied, and the different time behaviour is saved.

T: Manipulates the processor to compute the sensitive data using the target component (e.g., FPU).

O: Processor performs operations in the target hardware component using the sensitive data. It results in different execution time due to hardware optimizations.

R: Save the time to operate.

E: Analyze the saved data using the profile as a base.

**Attacks:** A benchmark is developed in [74] to measure the timing variability of floating point operations and report on its results. The approach uses floating point data timing variability to demonstrate practical attacks on the security of the Firefox browser (versions 23 through 27) and the Fuzz private database. In [75], an attack named LazyFPU that exploits x87 floating-point unit (FPU) is presented. This attack allows an adversary to recover the FPU and SIMD register set of arbitrary processes. The attack works on processors that transiently execute FPU or SIMD instructions that follow an instruction generating the fault indicating the first use of FPU or SIMD instructions.

*B. Memory Element*

In this section, various attacks which perform based on manipulation of the cache memory are described.

**Group DATA-MEM-MAN-I:** The first group refers to the Flush-and-Reload family, which relies on sharing pages between the attacker and the victim processes. With shared pages, the attacker can ensure that a specific memory line is evicted from the whole cache hierarchy. The spy uses this to monitor access to the memory line.

**Attack Formula:** In the following the phases of this attack has been presented:

- S: In the setup phase, the attacker gains memory access to monitor the memory, and the memory line is flushed from the cache hierarchy.
- T: After the setup, the attacker waits to allow the victim time to access the memory line before the next phase.
- O: This part represents an action that is done by the victim and leads the attacker to gain the desired information.
- R: In the Retrieve phase, the attacker reloads the memory line, together the information related to measuring the time to load it.
- E: In the Evaluation phase, the timing behavior is analyzed. If during the Trigger phase (waiting), the victim accesses the memory line which is available in the cache, thereload operation takes a short time. If, on the other hand, the victim has not accessed the memory line, the line needs to be brought from memory and the reload takes significantly longer. From this observation, the attacker can find out whether the victim has gained access to the cache or not.

**Attacks:** Flush+Relaod attack has been presented for the first time in [76]. In this work, the attack was studied on a single-core processor and exploits the adversary's process' ability to evict data in physical memory pages it shares with the victim process from the CPU cache (e.g., via the instruction clflush). In [77] another kind of Flush+Reload attack is presented. The difference is unlike most other attacks is able to make use of the side-channel information from almost all of the observed executions. This means it obtains private key recovery by observing a relatively small number of executions, approach in [78] uses the Flush+Reload attacks as a primitive, and extends it by leveraging within an automaton-driven strategy for tracing a victim's execution. This attack executed between tenants on commercial Platform-as-a-Service (PaaS) clouds. Work in [79] instead demonstrates Flush+Reload cache attacks in Virtual Machine. Finally, works in [20, 83–85] implemented Flush+Reload attacks in three steps and proposed a detection strategy for such attack.

**Group DATA-MEM-MAN-II:** There exist another cache attack technique called Flush+Flush. This attack relies only on the difference in timing of the flush instruction between cached and non cached memory accesses. In contrast to other cache attacks, it does not perform any memory accesses. Indeed, it builds upon the observation that the flush instruction leaks information on the state of the cache.

**Attack Formula:** In the following the phases of this attack has been presented:

- S: In the setup phase, the attacker gains access to memory to monitor the memory and the memory line is flushed from the cache hierarchy.
- T: After the setup, the attacker, waits to allow the victim time to access the memory line before the next phase.
- O: This part represents an action which is done by the victim and leads the attacker to gain the desirable information.
- R: In the Retrieve phase, the attacker flushes the memory line, together the information related to measuring the time to load it.
- E: In the Evaluation phase, the timing behavior is analyzed. The attacker measures the execution time of the flush instruction. Based on the execution time, the attacker decides whether the memory line has been cached or not. As the attacker does not load the memory line into the cache, this reveals whether some other process has loaded it. At the same time, clflush evicts the memory line from the cache for the next loop round of the attack. At the end of an attack round, the program optionally yields times in order to lower the system utilization and waits for the second process to perform some memory accesses.

**Attacks:** In [86] a Flush+Flush attack has been demonstrated. This attack is applicable in multi-core and virtualized environments if read-only shared memory with the victim process can be acquired. Authors in [85] performed Flush+Flush attack to an open source cache memory [97] and demonstrated a mitigation strategy for this attack based on address randomization.

**Group DATA-MEM-MAN-III:** EVICT+TIME and PRIME+PROBE are another attack which extracts the time measurement information by manipulating the state ofthe cache before each encryption, and observes the execution time of the subsequent encryption.

**Attack Formula:** In the following the phases of this attack has been presented:

- S: In the setup phase, the spy runs a spy process which monitors cache usage of the victim. Then, spy fills one or more cache sets with its own code or data (prime the cache).
- T: After the setup, spy waits while victim executes and utilizes the cache.
- O: This part represents an action which is done by the victim and leads the attacker to gain the desirable information.

- R: Retrieve: Spy continues execution (probe the cache) and measures the time to load each set of its data or code that is primed.
- E: In the Evaluation phase, the timing behavior is analyzed. If the victim has accessed some cache sets, it will have evicted some of the spy's lines, which spy observes as increased memory access latency for those lines. As the PROBE phase accesses the cache, it doubles as a PRIME phase for subsequent observations.

**Attacks:** One of the most general technique for Evict+Time and Prime+Probe are presented in [80]. It described their execution in a formal format. For instance in EVICT+TIME the attack extracts the time measurement information by manipulating the state of the cache before each encryption, and observes the execution time of the subsequent encryption. The article assumes the ability to trigger an encryption and know when it has begun and ended. It also assumes knowledge of the memory address of each lookup table, and hence of the cache sets to which it is mapped. In [81] the measurement method is the same as [80] Prime+Probe attack, but the analysis theory is based on the disalignment of AES lookup tables over L1 data cache and point out a way to detect the Lookup tables' residence in memory. [82] presents another Prime+Probe cache side-channel attack, which can prime physical addresses. These addresses are translated from virtual addresses used by a virtual machine. Then, time is measured to access these addresses, and it will be varied according to where the data is located. If it is in the CPU cache, the time will be less than in the main memory. The attack in this article was implemented in a server machine comparable to cloud environment servers. The technique in [83] is again based on Prime+Probe described in [80] but it is performed on the last level caches on virtual machine.

*C. Communication Element*

This part describes attacks to the communication structure aiming to retrieve sensitive data. Basically, these type of attacks targets on the latency or injection throughput to identify the features of a sensitive traffic. Examples of features are communication affinity (i.e., which nodes a specific node communicates with), communication data rates, size of the messages, etc. With this information, an attacker can model the communication behavior and infer the type of application is running, important nodes in the system, and the role of each node. However, there are different methods to accomplish these attacks, by creating the conditions (i.e., injection category), by manipulating the system to obtain the desirable conditions (i.e., manipulation category), or simply observing the normal system behavior (i.e., observation category). Therefore, there are three main groups of attacks, described in the following.

**Group DATA-COMM-INJ:** The first group refers to attacks where the message is crafted to force communication collisions. As a result of the collisions, the attacker can infer and extract sensitive information. In this group, we focus on attacks that modifies the priorities of the messages. The different priorities can change the behavior of the communication by creating back pressure on the buffers inside the routers. As a result, the attacker can do a fine-tuning to observe the sensitive traffic behavior.

**Attack Formula:** In the following the phases of this attack has been presented:

- S: In the setup phase, the attacker installs into a node, and start injecting packets using different priorities to understand the normal behavior.
- T: After the setup, the attacker choose the proper priority to create the expected collision.
- O: This part represents the system exchanging messages under different priorities. The target traffic is somehow affected or affects the attacker message.
- R: Retrieve: In the Retrieve phase, the attacker measures and analyze each injection to observe drops of performance. There is also an attack where the attacker observes a raise of performance when succeed.
- E: The different timing information collected is correlated with the setup used to attack. Based on delay to send packets, priority used, periodicity of target traffic, important features can be extracted.

**Attacks:** In the work [87], two attacks manipulating the priority of packets are described. The first one is called indirect congestion and forces collisions in the NoC when the attacker injects high priority packets. The second attack considers an attacker with no privilege to create high priority packets. In this case, the attacker forces a condition called backpressure, which basically observes when the injection of packets became faster due to the release of buffers in the path.

**Group DATA-COMM-MAN:** The second group refers to attacks that uses the communication structure as a mean to improve other architectural attacks that focus on Nodes or Memories. In this group, the attacker exploit the communication structure to identify the right instant to perform some malicious operation. As a result, it improves classical attacks by several orders.

**Attack Formula:** In the following the phases of this attack has been presented:

- S: In the setup phase, the attacker injects random packets (i.e., random size and random destinations) to understand the normal latency to inject a packet in the communication structure. Besides, in this phase, the attacker can install the malware in another node to observe other traffics that corresponds to the target Node or Memory.
- T: After the setup, the attacker asks for a service of the target node.
- O: The target node uses the communication structure to accomplish its tasks (e.g., retrieve data from a shared memory).
- R: In the Retrieve phase, the attacker injects messages in a way that collide with the target node paths.
- E: In the Evaluation phase, the different timing behavior show the presence of sensitive traffic. As a result, the attacker can infer in which part of the algorithm the target

Node is. Next, the attacker can reuse known methodologies to perform Node timing attacks or Memory access attacks.

**Attacks:** From this group, we can cite Earthquake [88], an attack that used the NoC to optimize the Differential Cache attack from Bogdanov. Results have shown that using Earthquake, the original cache attack became even more practical since the original required a high efficiency to work out. The works in [89] and [90] presented methodologies to use the NoC to improve cache access attacks like Prime+Probe. Finally, the work in [91] demonstrated that the same methodology used for NoCs can be used in bus-based systems as well.

**Group DATA-COMM-OBS:** The last group refers to pure communication timing attacks, where the attacker only observes the system behavior and retrieve its information.

**Attack Formula:** In the following the phases of this attack has been presented:

S: In the setup phase, the attacker injects random packets (i.e., random size and random destinations) to understand the normal latency to inject a packet in the communi- cation structure. Besides, in this phase, the attacker can install the malware in another node to observe other traffics.

T: After the setup, the attacker directly observes the traffic of the system.

O: The system nodes and memories use the communication structure to accomplish their tasks.

R: In the Retrieve phase, the attacker injects messages, that will collide with system traffic.

E: Evaluation: In the Evaluation phase, the latency to deliver a message is analyzed. If the time takes more than the normal measured during setup phase, the attacker consider a sensitive traffic. The amount of latency added, duration of this behavior and the instants in time can reveal important information like message size, data rate, and application type or behavior. The attacker can replace itself in the system to a different node, to obtain better efficiency to infer the system details.

**Attacks:** The works [93, 94] presented the first methodologies regarding the Network-on-Chip timing attacks. They showed how the injection latency could be explored as a leakage in the system. Later, the work in [92] detailed their methodologies by applying a practical use case. Additionally, the authors in [92] extended the NoC timing attack to a distributed timing attack, where other infected nodes could help to create controlled congestion in the network, which improved the observability of the attacker. In these attacks, the objective is to observe the communication behavior to infer sensitive traffic information.

## IX. DISCUSSIONS

In this section we would like to discuss the importance and benefits of having such taxonomy and attack model.

As described, the proposed taxonomy is created by exhaustively studying all different architectural attacks (we called it ArchA in this survey). Subsequently, we observed that all ArchA can be generally categorized based on three main metrics as follows: What, Where and How. What, basically refers to objective of the attacks which can be classified as *obtain intellectual property information*, *modify the functionality* and, *steal data*. Where instead, refers to locations that attackers perform the attacks. In this survey these locations are identified as *Processing elements*, *Memories* and *Communication components*. How, describes the method by which attackers execute attacks. These methods are considered as *Injection*, *Manipulation* and *Observation*.

The proposed taxonomy exceeds the state-of-the-art in several aspects as follows i) by analysing all different objectives, in the contrary to the work in [8] which is only focused on attacks which steal data. ii) by taking to account different possible locations, in contrary to approaches in [9, 10, 13] which only are focused on caches iii) by introducing a new aspect which classify ArchA based on the method they execute the attacks. To the best of authors acknowledge this classification is introduced by the first time in this work.

On the other hand this survey introduced an innovative attack model called STORE which identifies five main phases for all ArchA to be executed. Accordingly, STORE is demon- strated as follows: a Setup phase for preparation of attacks, Trigger phase regarding actions to activate vulnerability, Op- eration phase for the execution and exposing vulnerabilities, Retrieve phase regarding collecting essential information for attacks, and finally Evaluation phase for analyzing the gathered information to obtain the final attack's objective.

The proposed attack model is an essential since same attacks in the stat-of-the-art are introduced with different amount of phases. This means that any authors have a different view for analyzing attacks. For instance, work in [17] divided Spectre attack to three steps while approach in [22] and [18] categorized it in two and five phases, respectively. The other example is that, authors in [14] and [98] divided FLUSH+RELOAD in three phases, however the definition of these phases are completely differ from each other.

However, STORE analyzed all the existing ArchA from one point of view and in consistent phases.

All in all having such a complete and unified taxonomy and attack model leads to an easier elaborating countermeasures. Moreover, it contributes to developing more accurate security tools and prediction of possible attacks.

## X. CONCLUSIONS

We presented a novel taxonomy for ArchA which uses three different metrics, i.e., What, Where and How, to analyze each attack type. These metrics together with the proposed taxonomy outperform the previous ones since they successfully categorized all the existed ArchA in clear and unified groups. We furthermore introduced an appropriate attack model that uses five different phases i.e. Setup, Trigger, Operation, Retrieve and Evaluation, to formally describe sequence of actions in any ArchA. Similar to the taxonomy problem most attack models only cover a subset of attacks and they are mainly focused on caches. In addition, we want to stress that having such a clear taxonomy and a unified attack model leads to a better understating of the attacks' behavior, thus, facilitating development of a system which is able to mitigate such attacks.

## REFERENCES

[1] Fortune Business Insights, "Cyber security market size." [Online]. Available: https://www.fortunebusinessinsights.com/industry-reports/cyber-security-market-101165
[2] "5g security and resilience published by Cybersecurity and Infrastructure security agency, url = https://www.cisa.gov/5g, urldate = 2021-09-12,."
[3] A. A. Cárdenas, T. Roosta, G. Taban, and S. Sastry, "Cyber security basic defenses and attack trends," *Homeland Security Technology Challenges*, pp. 73–101, 2008.
[4] J. Jang-Jaccard and S. Nepal, "A survey of emergingthreats in cybersecurity," *Journal of Computer and System Sciences*, vol. 80, no. 5, pp. 973–993, 2014, special Issueon Dependable and Secure Computing. [Online]. Available: https://www.sciencedirect.com/science/article/pii/S0022000014000178
[5] V. Tiwari and Dwivedi, "Analysis of cyber attack vectors," 04 2016.
[6] Y. Kim, W. C. Lau, M. C. Chuah, and H. Chao, "Packetscore: a statistics-based packet filtering scheme against distributed denial-of-service attacks," *IEEE Transactions on Dependable and Secure Computing*, vol. 3, no. 2, pp. 141–155, 2006.
[7] H. Wang, H. Li, F. Rahman, M. M. Tehranipoor, and F. Farahmandi, "Sofi: Security property-driven vulnerability assessments of ics against fault-injection attacks," *IEEE Transactions on Computer-Aided Design of Integrated Circuits and Systems*, pp. 1–1, 2021.
[8] A. Bogdanov, T. Eisenbarth, C. Paar, and M. Wienecke, "Differential cache-collision timing attacks on aes with applications to embedded cpus," in *CT-RSA 2010*.
[9] T. Zhang and R. Lee, "Secure cache modeling for measuring side-channel leakage," *Technical Report, Princeton University*, 2014.
[10] A. Canteaut, C. Lauradoux, and A. Seznec, "Understanding cache attacks," INRIA, Research Report RR-5881, 2006. [Online]. Available: https://hal.inria.fr/inria-00071387
[11] P. Kocher, D. Genkin, D. Gruss, W. Haas, M. Hamburg, M. Lipp, S. Mangard, T. Prescher, M. Schwarz, and Y. Yarom, "Spectre attacks: Exploiting speculative execution," *arXiv preprint arXiv:1801.01203*, 2018.
[12] Y. Lyu and P. Mishra, "A survey of side-channel attacks on caches and countermeasures," *Journal of Hardware and Systems Security*, vol. 2, no. 1, pp. 33–50, Mar 2018.
[13] Q. Ge, Y. Yarom, D. Cock, and G. Heiser, "A survey of microarchitectural timing attacks and countermeasures on contemporary hardware," *Journal of Cryptographic Engineering*, vol. 8, no. 1, pp. 1–27, 2018.
[14] S. Deng, W. Xiong, and J. Szefer, "Analysis of secure caches using a three-step model for timing-based attacks," Cryptology ePrint Archive, Report 2019/167, 2019, https://eprint.iacr.org/2019/167.
[15] Y. Yarom and K. Falkner, "Flush+reload: A high resolution, low noise, l3 cache side-channel attack," in *23rd USENIX*, 2014.
[16] D. A. Osvik, A. Shamir, and E. Tromer, "Cache attacks and countermeasures: The case of AES," in *CT-RSA 2006,San Jose, CA, USA, February 13-17, 2006*, 2006.
[17] P. Kocher and et al., "Spectre Attacks: Exploiting Speculative Execution," in *IEEE SP*, 2019.
[18] C. Canella, J. Van Bulck, M. Schwarz, M. Lipp, B. Von Berg, P. Ortner, F. Piessens, D. Evtyushkin, and D. Gruss, "A systematic evaluation of transient execution attacks and defenses," in *Proceedings of the 28th USENIX Conference on Security Symposium*, ser. SEC'19, 2019.
[19] D. Budgen and P. Brereton, "Performing systematic literature reviews in software engineering," in *Proceedings of the 28th International Conference on Software Engineering*, ser. ICSE '06. New York, NY, USA: Association for Computing Machinery, 2006, p. 1051–1052. [Online]. Available: https://doi.org/10.1145/1134285.1134500
[20] C. Reinbrecht, S. Hamdioui, M. Taouil, B. Niazmand, T. Ghasempouri, J. Raik, and J. Sepúlveda, "Lid-cat: A lightweight detector for cache attacks," in *2020 IEEE European Test Symposium (ETS)*, 2020, pp. 1–6.
[21] D. J. Bernstein, "Cache timing attacks on aes," April 2005, accessed: 2016-12-31.
[22] J. Bonneau and I. Mironov, "Cache-Collision Timing Attacks Against AES," in *CHES*, Yokohama, Japan, October 2006, pp. 201–215.
[23] C. Domas, "Breaking the x86 isa," 2017.
[24] P. Koppe, B. Kollenda, M. Fyrbiak, C. Kison, R. Gawlik, C. Paar, and T. Holz, "Reverse engineering x86 processor microcode," in *Proceedings of the 26th USENIX Conference on Security Symposium*, ser. SEC'17, 2017, p. 1163–1180.
[25] H. Wong, M.-M. Papadopoulou, M. Sadooghi-Alvandi, and A. Moshovos, "Demystifying gpu microarchitecture through microbenchmarking," in *2010 IEEE International Symposium on Performance Analysis of Systems Software (ISPASS)*, 2010, pp. 235–246.
[26] C. Domas, "God mode unlocked: Hardware backdoors in x86 cpus," 2018.
[27] J. Dongarra, S. Moore, P. Mucci, K. Seymour, and H. You, "Accurate cache and tlb characterization using hardware counters," in *Computational Science - ICCS 2004*, M. Bubak, G. D. van Albada, P. M. A. Sloot, and J. Dongarra, Eds. Berlin, Heidelberg: Springer Berlin Heidelberg, 2004, pp. 432–439.
[28] C. Coleman and J. Davidson, "Automatic memory hierarchy characterization," in *2001 IEEE International Symposium on Performance Analysis of Systems and Software. ISPASS.*, 2001, pp. 103–110.
[29] M. Jung, C. C. Rheinländer, C. Weis, and N. Wehn, "Reverse engineering of drams: Row hammer with crosshair," in *Proceedings of the Second International Symposium on Memory Systems*. New York, NY, USA: Association for Computing Machinery, 2016, p. 471–476.
[30] M. Hassan, A. M. Kaushik, and H. Patel, "Exposing implementation details of embedded dram memory controllers through latency-based analysis," *ACM Trans. Embed. Comput. Syst.*, vol. 17, no. 5, Oct. 2018. [Online]. Available: https://doi.org/10.1145/3274281
[31] A. Abel and J. Reineke, "Measurement-based modeling of the cache replacement policy," in *2013 IEEE 19th Real-Time and Embedded Technology and Applications Symposium (RTAS)*, 2013, pp. 65–74.
[32] M. Hassan, A. M. Kaushik, and H. Patel, "Reverse-engineering embedded memory controllers through latency-based analysis," in *21st IEEE Real-Time and Embedded Technology and Applications Symposium*, 2015, pp. 297–306.
[33] S. Evain and J.-P. Diguet, "From noc security analysis to design solutions," in *IEEE Workshop on Signal Processing Systems Design and Implementation, 2005.*, 2005, pp. 166–171.
[34] A. Baratloo, N. Singh, T. K. Tsai *et al.*, "Transparent run-time defense against stack-smashing attacks." in *USENIX Annual Technical Conference, General Track*, 2000, pp. 251–262.
[35] G. Richarte *et al.*, "Four different tricks to bypass stackshield and stackguard protection," *World Wide Web*, vol. 1, 2002.
[36] S. Bhatkar, D. C. DuVarney, and R. Sekar, "Address obfuscation: An efficient approach to combat a broad range of memory error exploits." in *USENIX Security Symposium*, vol. 12, no. 2, 2003, pp. 291–301.
[37] H. Shacham, M. Page, B. Pfaff, E.-J. Goh, N. Modadugu, and D. Boneh, "On the effectiveness of address-space randomization," in *Proceedings of the 11th ACM conference on Computer and communications security*. ACM, 2004, pp. 298–307.
[38] A. N. Sovarel, D. Evans, and N. Paul, "Where's the feeb? the effectiveness of instruction set randomization." in *USENIX Security Symposium*, vol. 10, 2005.
[39] S. Alexander, "Defeating compiler-level buffer overflow protection," *The USENIX Magazine; login*, 2005.
[40] H. Ozdoganoglu, T. Vijaykumar, C. E. Brodley, B. A. Kuperman, and A. Jalote, "Smashguard: A hardware solution to prevent security attacks on the function return address," *IEEE Transactions on Computers*, vol. 55, no. 10, pp. 1271–1285, 2006.
[41] Y. Lee, I. Heo, D. Hwang, K. Kim, and Y. Paek, "Towards a practical solution to detect code reuse attacks on arm mobile devices," in *Proceedings of the Fourth Workshop on Hardware and Architectural Support for Security and Privacy*. ACM, 2015, p. 3.
[42] N. P. Smith, "Stack smashing vulnerabilities in the unix operating system," 1997.
[43] C. Cowan, C. Pu, D. Maier, J. Walpole, P. Bakke, S. Beattie, A. Grier, P. Wagle, Q. Zhang, and H. Hinton, "Stackguard: Automatic adaptive detection and prevention of buffer-overflow attacks." in *USENIX Security Symposium*, vol. 98. San Antonio, TX, 1998, pp. 63–78.
[44] O. Aleph, "Smashing the stack for fun and profit," *http://www.shmoo.com/phrack/Phrack49/p49-14*, 1996.

[45] N. Carlini, A. Barresi, M. Payer, D. Wagner, and T. R. Gross, "Control-flow bending: On the effectiveness of control-flow integrity." in *USENIX Security Symposium*, 2015, pp. 161–176.

[46] D. Gruss, M. Lipp, M. Schwarz, R. Fellner, C. Maurice, and S. Mangard, "Kaslr is dead: long live kaslr," in *International Symposium on Engineering Secure Software and Systems*. Springer, 2017, pp. 161–176.

[47] Y. Jang, J. Lee, S. Lee, and T. Kim, "Sgx-bomb: Locking down the processor via rowhammer attack," in *Proceedings of the 2nd Workshop on System Software for Trusted Execution*. ACM, 2017, p. 5.

[48] V. Van Der Veen, Y. Fratantonio, M. Lindorfer, D. Gruss, C. Maurice, G. Vigna, H. Bos, K. Razavi, and C. Giuffrida, "Drammer: Deterministic rowhammer attacks on mobile platforms," in *Proceedings of the 2016 ACM SIGSAC conference on computer and communications security*. ACM, 2016, pp. 1675–1689.

[49] P. Pessl, D. Gruss, C. Maurice, M. Schwarz, and S. Mangard, "Drama: Exploiting dram addressing for cross-cpu attacks." in *USENIX Security Symposium*, 2016, pp. 565–581.

[50] Y. Kim, R. Daly, J. Kim, C. Fallin, J. H. Lee, D. Lee, C. Wilkerson, K. Lai, and O. Mutlu, "Flipping bits in memory without accessing them: An experimental study of dram disturbance errors," in *ACM SIGARCH Computer Architecture News*, vol. 42, no. 3. IEEE Press, 2014, pp. 361–372.

[51] D. Gruss, C. Maurice, and S. Mangard, "Rowhammer. js: A remote software-induced fault attack in javascript," in *International Conference on Detection of Intrusions and Malware, and Vulnerability Assessment*. Springer, 2016, pp. 300–321.

[52] A. Tatar, R. Krishnan, E. Athanasopoulos, C. Giuffrida, H. Bos, and K. Razavi, "Throwhammer: Rowhammer attacks over the network and defenses," in *2018 USENIX Annual Technical Conference*. USENIX Association, 2018.

[53] M. Lipp, M. T. Aga, M. Schwarz, D. Gruss, C. Maurice, L. Raab, and L. Lamster, "Nethammer: Inducing rowhammer faults through network requests," *arXiv preprint arXiv:1805.04956*, 2018.

[54] D. Grunwald and S. Ghiasi, "Microarchitectural denial of service: Insuring microarchitectural fairness," in *Proceedings of the 35th Annual ACM/IEEE International Symposium on Microarchitecture*, ser. MICRO 35. Los Alamitos, CA, USA: IEEE Computer Society Press, 2002, pp. 409–418.

[55] T. Boraten and A. K. Kodi, "Mitigation of denial of service attack with hardware trojans in noc architectures," in *2016 IEEE International Parallel and Distributed Processing Symposium (IPDPS)*, May 2016, pp. 1091–1100.

[56] J. Sepúlveda, M. Strum, and W. Chau, "An hybrid switching approach for noc-based systems to avoid denial-of-service soc attacks," *16th Iberchip Wksp (IWS 2010)*, pp. 23–25, 2010.

[57] L. L. Caimi, V. Fochi, E. Wachter, D. Munhoz, and F. G. Moraes, "Activation of secure zones in many-core systems with dynamic rerouting," in *2017 IEEE International Symposium on Circuits and Systems (ISCAS)*, May 2017, pp. 1–4.

[58] L. Fiorin, G. Palermo, and C. Silvano, "A security monitoring service for nocs," in *Proceedings of the 6th IEEE/ACM/IFIP International Conference on Hardware/Software Codesign and System Synthesis*, ser. CODES+ISSS '08. New York, NY, USA: ACM, 2008, pp. 197–202.

[59] J. Sepúlveda, D. Flórez, and G. Gogniat, "Efficient and flexible noc-based group communication for secure mpsocs," in *2015 International Conference on ReConFigurable Computing and FPGAs (ReConFig)*, Dec 2015, pp. 1–6.

[60] S. Evain and J. P. Diguet, "From noc security analysis to design solutions," in *IEEE Workshop on Signal Processing Systems Design and Implementation, 2005.*, Nov 2005, pp. 166–171.

[61] R. JS, D. M. Ancajas, K. Chakraborty, and S. Roy, "Runtime detection of a bandwidth denial attack from a rogue network-on-chip," in *Proceedings of the 9th International Symposium on Networks-on-Chip*. ACM, 2015, p. 8.

[62] L. Fiorin, C. Silvano, and M. Sami, "Security aspects in networks-on-chips: Overview and proposals for secure implementations," in *Digital System Design Architectures, Methods and Tools, 2007. DSD 2007. 10th Euromicro Conference on*. IEEE, 2007, pp. 539–542.

[63] A. B. Achballah, S. B. Othman, and S. B. Saoud, "Toward on hardware firewalling of networks-on-chip based systems," in *2017 International Conference on Advanced Systems and Electric Technologies (IC_ASET)*, Jan 2017, pp. 7–13.

[64] M. D. Grammatikakis, K. Papadimitriou, P. Petrakis, A. Papagrigoriou, G. Kornaros, I. Christoforakis, O. Tomoutzoglou, G. Tsamis, and M. Coppola, "Security in mpsocs: A noc firewall and an evaluation framework," *IEEE Transactions on Computer-Aided Design of Integrated Circuits and Systems*, vol. 34, no. 8, pp. 1344–1357, Aug 2015.

[65] J. Van Bulck, F. Piessens, and R. Strackx, "Nemesis: Studying microarchitectural timing leaks in rudimentary cpu interrupt logic," in *Proceedings of the 2018 ACM SIGSAC Conference on Computer and Communications Security*, ser. CCS '18. New York, NY, USA: ACM, 2018, pp. 178–195.

[66] M. Schwarz, M. Schwarzl, M. Lipp, and D. Gruss, "Netspectre: Read arbitrary memory over network," *arXiv preprint arXiv:1807.10535*, 2018.

[67] S. van Schaik, A. Milburn, S. Österlund, P. Frigo, G. Maisuradze, K. Razavi, H. Bos, and C. Giuffrida, "RIDL: Rogue in-flight data load," in *S&P*, May 2019.

[68] M. Minkin, D. Moghimi, M. Lipp, M. Schwarz, J. Van Bulck, D. Genkin, D. Gruss, B. Sunar, F. Piessens, and Y. Yarom, "Fallout: Reading kernel writes from user space," 2019.

[69] M. Lipp, M. Schwarz, D. Gruss, T. Prescher, W. Haas, S. Mangard, P. Kocher, D. Genkin, Y. Yarom, and M. Hamburg, "Meltdown," *arXiv preprint arXiv:1801.01207*, 2018.

[70] J. V. Bulck, M. Minkin, O. Weisse, D. Genkin, B. Kasikci, F. Piessens, M. Silberstein, T. F. Wenisch, Y. Yarom, and R. Strackx, "Foreshadow: Extracting the keys to the intel SGX kingdom with transient out-of-order execution," in *27th USENIX Security Symposium (USENIX Security 18)*. Baltimore, MD: USENIX Association, Aug. 2018, p. 991–1008. [Online]. Available: https://www.usenix.org/conference/usenixsecurity18/presentation/bulck

[71] D. Evtyushkin, R. Riley, N. C. Abu-Ghazaleh, ECE, and D. Ponomarev, "Branchscope: A new side-channel attack on directional branch predictor," in *Proceedings of the Twenty-Third International Conference on Architectural Support for Programming Languages and Operating Systems*, ser. ASPLOS '18. New York, NY, USA: ACM, 2018, pp. 693–707. [Online]. Available: http://doi.acm.org/10.1145/3173162.3173204

[72] G. Maisuradze and C. Rossow, "Speculose: Analyzing the security implications of speculative execution in cpus," *CoRR*, vol. abs/1801.04084, 2018. [Online]. Available: http://arxiv.org/abs/1801.04084

[73] E. M. Koruyeh, K. N. Khasawneh, C. Song, and N. Abu-Ghazaleh, "Spectre returns! speculation attacks using the return stack buffer," in *12th USENIX Workshop on Offensive Technologies ( WOOT 18)*, 2018.

[74] M. Andrysco, D. Kohlbrenner, K. Mowery, R. Jhala, S. Lerner, and H. Shacham, "On subnormal floating point and abnormal timing," in *Security and Privacy (SP), 2015 IEEE Symposium on*. IEEE, 2015, pp. 623–639.

[75] J. Stecklina and T. Prescher, "Lazyfp: Leaking fpu register state using microarchitectural side-channels," *arXiv preprint arXiv:1806.07480*, 2018.

[76] D. Gullasch, E. Bangerter, and S. Krenn, "Cache Games - Bringing Access-Based Cache Attacks on AES to Practice," in *IEEE SP*, May 2011, pp. 490–505.

[77] N. Benger, J. Pol, N. P. Smart, and Y. Yarom, "Ooh Aah... Just a Little Bit: A Small Amount of Side Channel Can Go a Long Way," in *CHES*, New York, NY, USA, 2014, pp. 75–92.

[78] Y. Zhang, A. Juels, M. K. Reiter, and T. Ristenpart, "Cross-Tenant Side-Channel Attacks in PaaS Clouds," in *ACM SIGSAC*, 2014, pp. 990–1003.

[79] G. Irazoqui, M. S. Inci, T. Eisenbarth, and B. Sunar, *Wait a Minute! A fast, Cross-VM Attack on AES*. Springer International Publishing, 2014, pp. 299–319.

[80] D. Osvik and et al., "Cache attacks and countermeasures: The case of aes," in *Topics in Cryptology - CT-RSA 2006*, 2006, pp. 1–20.

[81] Z. Xinjie and et al., "Robust first two rounds access driven cache timing attack on aes," in *CSSE*, vol. 3, Hubei, China, Dec 2008, pp. 785–788.

[82] Y. A. Younis and et al., "A new prime and probe cache side-channel attack for cloud computing," in *CSIT*, Oct 2015, pp. 1718–1724.

[83] T. Ghasempouri, J. Raik, K. Paul, C. Reinbrecht, S. Hamdioui, and M. Taouil, "A security verification template to assess cache architecture vulnerabilities," in *2020 23rd International Symposium on Design and Diagnostics of Electronic Circuits Systems (DDECS)*, 2020, pp. 1–6.

[84] T. Ghasempouri, J. Raik, K. Paul, C. Reinbrecht, S. Hamdioui,and Mottaqiallah, "Verifying cache architecture vulnerabilities using a formal security verification flow," *Microelectronics*

*Reliability*, vol. 119, p. 114085, 2021. [Online]. Available: https://www.sciencedirect.com/science/article/pii/S0026271421000512

[85] A. Shalabi, T. Ghasempouri, P. Ellervee, and J. Raik, "Scaat: Secure cache alternative address table for mitigating cache logical side-channel attacks," in *2020 23rd Euromicro Conference on Digital System Design (DSD)*, 2020, pp. 213–217.

[86] F. Liu, Y. Yarom, Q. Ge, G. Heiser, and R. B. Lee, "Last-Level Cache Side-Channel Attacks are Practical," in *IEEE SP*, San Jose, CA, USA, 2015, pp. 605–622.

[87] B. Forlin, C. Reinbrecht, and J. Sepúlveda, "Attacking real-time mpsocs: Preemptive nocs are vulnerable," in *2019 IFIP/IEEE 27th International Conference on Very Large Scale Integration (VLSI-SoC)*, 2019, pp. 204–209.

[88] C. Reinbrecht, B. Forlin, A. Zankl, and J. Sepúlveda, "Earthquake — a noc-based optimized differential cache-collision attack for mpsocs," in *2018 Design, Automation Test in Europe Conference Exhibition (DATE)*, March 2018, pp. 648–653.

[89] "Timing attack on NoC-based systems: Prime+Probe attack and NoC-based protection," *Microprocessors and Microsystems*, vol. 52, no. Supplement C, pp. 556 – 565, 2017.

[90] C. Reinbrecht, A. Susin, L. Bossuet, G. Sigl, and J. Sepúlveda, "Side channel attack on noc-based mpsocs are practical: Noc prime+probe attack," in *2016 29th Symposium on Integrated Circuits and Systems Design (SBCCI)*, Aug 2016, pp. 1–6.

[91] J. Sepúlveda, M. Gross, A. Zankl, and G. Sigl, "Exploiting bus communication to improve cache attacks on systems-on-chips," in *2017 IEEE Computer Society Annual Symposium on VLSI (ISVLSI)*, July 2017, pp. 284–289.

[92] C. Reinbrecht, A. Susin, L. Bossuet, and J. Sepulveda, "Gossip NoC - Avoiding Timing Side-Channel Attacks through Traffic Management," in *ISVLSI 16*. Pittsburgh, USA: IEEE, July 2016, pp. 601–606.

[93] J. Sepulveda, J. Diguet, M. Strum, and G. Gogniat, "NoC-Based Protection for SoC Time-Driven Attacks," *Embedded Systems Letters, IEEE*, vol. 7, no. 1, pp. 7–10, March 2015.

[94] W. Yao and E. Suh, "Efficient timing channel protection for on-chip networks," in *NOCS*, Lyngby, Denmark, May 2012, pp. 142–151.

[95] Software Engineering Institute, Carnegie Mellon University, "Cert coordination center." [Online]. Available: https://www.sei.cmu.edu/about/divisions/cert/index.cfm

[96] Y. Ruan, S. Kalyanasundaram, and X. Zou, "Survey of return-oriented programming defense mechanisms," *Security and Communication Networks*, vol. 9, no. 10, pp. 1247–1265, 2016.

[97] Chair of VLSI Design, Diagnostics and Architecture. (2016) PoC -Pile of Cores. Technische Universität Dresden. [Online]. Available: https://github.com/VLSI-EDA/PoC

[98] D. Gruss, C. Maurice, K. Wagner, and S. Mangard, "Flush+flush: A fast and stealthy cache attack," in *13th International Conference on Detection of Intrusions and Malware, and Vulnerability Assessment*, 2016.